\documentstyle[12pt,epsfig]{article}

\newcommand{\f}{\begin{equation}}
\newcommand{\ff}{\end{equation}}
\newcommand{\beq}{\begin{equation}}
\newcommand{\eeq}{\end{equation}}
\newcommand{\bea}{\begin{eqnarray}}
\newcommand{\eea}{\end{eqnarray}}

\newcommand{\C}{{\cal C}}
\newcommand{\Screen}{S}

\newcommand{\CP}{\mbox{\sl P}}
\newcommand{\NP}{\mbox{\sl NP}}
\newcommand{\LNP}{\mbox{\sl LNP}}
\newcommand{\RNP}{\mbox{\sl RNP}}
\newcommand{\TP}{\mbox{\sl TP}}
\newcommand{\F}{\mbox{\sl F}}
\newcommand{\NF}{\mbox{\sl NF}}
\newcommand{\ES}{\mbox{\sl ES}}
\newcommand{\TF}{\mbox{\sl TF}}
\newcommand{\LR}{\mbox{\sl LR}}
\newcommand{\RL}{\mbox{\sl RL}}

\newcommand{\QSN}{\mbox{\sl QS}}
\newcommand{\dimension}{\mbox{\rm dim}}

\makeatletter
\newcommand{\ps@preprint}{%
        \renewcommand{\@oddhead}{\hfil\small{CGPG-99/10}}}%
\makeatother
\begin{document}
\title{Holography in a quantum spacetime}
\author{Fotini Markopoulou\thanks{fotini@phys.psu.edu}\
        and Lee Smolin\thanks{smolin@phys.psu.edu}\\
	 \\
	 	Center for Gravitational Physics and Geometry\\
	Department of Physics\\
	The Pennsylvania State University\\
	University Park, PA 16802, USA
	 \\
	 and
	 \\
	 The Blackett Laboratory\\
	 Imperial College of Science, Technology and Medicine\\
	 South Kensington\\
	 London SW7 2BZ, UK\\}

\date{October 18, 1999}  
\maketitle
\thispagestyle{preprint}  
\vfill
\begin{abstract}
We propose a formulation of the holographic principle, suitable for a 
background independent quantum theory of cosmology. It is stated as a 
relationship between the flow of quantum information and the causal 
structure of a quantum spacetime.  Screens are defined 
as sets of events at which the observables of a holographic cosmological 
theory may be measured, and such that information may
flow across them in two directions. A discrete background independent 
holographic theory may be formulated in terms of information flowing 
in a causal network of such screens.  Geometry is introduced by {\it defining}
the area of a screen to be a measure of its  capacity 
as a channel of quantum information from its null past to its null future.
We call this a ``weak'' form  of the holographic principle, as  no
use is made of a bulk theory. 
 
\end{abstract}
\vfill
\newpage

\section{Introduction}

In this paper, we present a framework for a Planck-scale, 
cosmological, background-independent theory which is holographic in 
a sense appropriate to a quantum spacetime.  This is motivated by the 
fact that the formulations of the holographic principle given to 
date\cite{thooft-holo}-\cite{others}
are confined to the semiclassical regime.  At the same time, results 
of several approaches to quantum gravity indicate that the description
of spacetime based on smooth manifolds  can provide only an approximate 
description\cite{roger-sn}-\cite{strings-finite}.  
If it is true, the holographic principle ought then to be 
more than a conjecture about the classical and semiclassical theory.  
Rather, it should be an important part of the framework of a 
Planck-scale, background independent quantum theory. 

Our goal is to give a form of the holographic 
principle that could be satisfied by a background independent
quantum theory of gravity, and reduces to the standard 
holographic principle  in the semiclassical limit.  To guide us,
we make  two assumptions.  First, the theory must be cosmological,
in the sense that whatever structure replaces the smooth spacetime 
geometry
will have no fixed
external boundary or asymptotic regions. This means that we must
keep in mind what has been learned about how to formulate the
holographic principle at the semiclassical level in cosmological
spacetimes.  One important lesson, discovered first by Fischler and
Susskind\cite{willylenny} and developed by Bousso\cite{RB}, is that 
in a cosmological
spacetime the holographic principle must be formulated in a way that
employs the lightcones of the spacetime.  Otherwise,
paradoxes arise which are discussed in \cite{willylenny,RB,others}.
Some of these paradoxes concern cases in which the number of bulk degrees of
freedom associated with non-null surfaces fails to be bounded,
or even fails to be well defined.
These same problems arise when we attempt to 
formulate a version of the holographic principle suitable for a
background independent theory\cite{leeholo}.   
To avoid them, we assume here that there is
a fundamental causal structure which plays a role in the background
independent theory and that in the classical limit this becomes
the causal structure of a spacetime. This is our second assumption.  
 
To proceed, we need a framework for a cosmological quantum
theory which incorporates causal structure. We require that this 
framework be sufficiently general that it can accomodate a 
background independent quantum  theory of gravity.
Such a framework, called quantum causal histories, was  
defined in \cite{FM4}.  It is cosmological  in the sense 
that any physically meaningful observable corresponds to some observer 
(represented as an event or collection of events) inside the closed universe. 
(In the classical case, see \cite{FM3}).  As each 
observer receives information from a distinct past, the algebra of 
observables they can measure, and hence the (finite-dimensional) 
Hilbert spaces on which 
what they observe can be represented, vary over the history.  
Consequently, the algebra of observables of the theory is represented on a 
collection of Hilbert spaces. These replace the single wavefunction
and single Hilbert space of other approaches to quantum cosmology.

Quantum causal histories were originally motivated by the need to
provide a general framework to understand what observables are in
background independent approaches to quantum gravity, such as 
those proposed in \cite{FM2,tubes,pqtubes,mpaper}.
In these theories a causal quantum spacetime is constructed from local
changes in a spin network or a network of abstract surfaces.  The Hilbert
spaces used to describe such changes are finite-dimensional because
only a finite amount of information about the quantum geometry is
involved in each elementary causal process.
As one expects from a scheme of this type, in which   a fundamental 
discreteness  in the structure of spacetime appears at Planck scales, 
a key question  the theory must answer is whether a smooth spacetime
geometry is recovered in the continuum limit.  

The question we address in this paper is what properties such a
theory must have in order that its semiclassical limits, when they
exist, are holographic in the sense defined in 
\cite{thooft-holo,lenny-holo,willylenny,RB,others}.  
A central element in the   
semiclassical formulations of 
the cosmological holographic principle is that of a screen, 
a spacelike 2-surface
on which the relevant degrees of freedom of the theory live\footnote{
In dimensions other than $4$, a screen is a 
spacelike surface of codimension $2$ in the spacetime.}\cite{RB}.  We 
require that there are analogues of these 2-surfaces in the quantum 
spacetime.  In the next section we define what we call an ``elementary
screen'', these are certain collections of events in a quantum
causal history.  We then define a class of quantum
spacetimes consisting of causal 
networks of such screens.  We call these {\it screen networks}.  We show
in section 4 that there are examples of them which may be constructed
by imposing certain restrictions on the class of background
independent theories of quantum gravity and string theory given
in \cite{FM2,tubes,pqtubes,mpaper}.

An obvious fact about 
2-dimensional surfaces turns out to be key in this work, namely, that 
each 2-surface has two sides.  We find below that this has
two important consequences.  First, the additional structure in a screen 
network which follows from the two-sided nature of the elementary screens allows 
a distinction between null and timelike propagation, something that an 
ordinary causal set history does not provide.  Second, this makes it possible 
to incorporate chirally asymmetric theories in the causal quantum 
history framework. 

An essential element of the holographic principle in its semiclassical 
forms is the Bekenstein bound\cite{bek}.   This must be recovered in the 
semiclassical limit of any holographic theory.  This seems to present 
a potential problem, as metric plays no role in the definition of a 
screen network.  Given that the metric is unified with
other degrees of freedom in string theory, it is not even clear
that the notion of area should be well defined at the background
independent level.  The only property a screen has
beyond its place in the causal network is the dimension of its
Hilbert space.   This leads us to suggest that  
the Bekenstein bound may be 
inverted and {\it area be defined to be a measure of the capacity of a 
screen for the transmission of quantum information}.  

The result is a form of the holographic principle which makes no use
at all of the notion of a bulk theory, and instead posits only a
relationship between the information capacity and the geometrical 
area of a screen.  We call this a ``weak holographic principle'', and 
reserve ``strong'' for those formulations of the principle that posit 
relationships between bulk and boundary theories, or limits on the 
amount of information on spacelike or null surfaces bounded by the 
screens.  We believe that strong forms of the principle are only 
relevant in the semiclassical theory, when the bound is on the number 
of matter degrees of freedom in a region of a single fixed spacetime and that, 
once the gravitational degrees of freedom are introduced, either 
classically or quantum mechanically, only weak forms of the principle 
are possible.

A key feature of the weak holographic principle is that 
a complete description of the universe requires more than one screen.  
This is simply because in a generic cosmological history there is no 
single screen whose past is the entire universe.  Thus, a 
cosmological holographic theory must be a many-screens theory, each screen 
recording information about its causal past.  It is important to note 
that such a many-screens theory gives us the possibility to dispense 
with the notion of the bulk theory.  Rather than formulate the 
holographic principle in terms of a relationship  between a bulk 
theory and a boundary, as is done in its strong forms, we can 
formulate it entirely in terms of screen observables and relationships 
between them.

A question left open for future work is the exact relationship
between weak and strong forms of the holographic principle.
It may be conjectured that when the weak holographic 
principle applies to a quantum causal history which has a good 
classical 
limit, the strong holographic principle holds in that limit.  
We expect that, when the limit exists,  
2-dimensional surfaces in the continuum spacetime will
originate from ensembles of elementary screens and that their area 
will satisfy the Bekenstein bound in the standard way.  At the same 
time, we expect that no cosmological form of the holographic
principle, even one that holds in the semiclassical limit, can
escape the fact that many screens are necessary to give a
complete description of a cosmological spacetime.
 
The outline of this paper is as follows.  In the next section we 
define screens and screen networks.    In section 3, a quantum screen 
network is defined, following the general prescription of \cite{FM4}, 
as a functor from the edge-sets of a screen network to the Hilbert space 
category.  In section 4 we show what restrictions are imposed on
the background independent formulations of quantum gravity and
string theory given in \cite{FM2,tubes,pqtubes,mpaper} if one requires
that all events are elementary screens.  
In section 4 the essential features of this
approach to the holographic principle are then abstracted and 
formulated as a proposal for the {\it weak holographic principle.}
Comments on the correspondence to the semiclassical 
holographic principle as well as on the conditions required, were a
strong form of the holographic principle to hold in a background
independent theory, can be found in 
the final section. 
 
\section{Elementary screens and screen networks}
 
We begin by recalling the definition of a causal set, as used
by Sorkin, 't Hooft and others to describe the discrete analogue
of the causal relations of events in a Lorentzian spacetime 
(see \cite{blms,sorkin,dm,tHooft}).  Then we explain what causal histories are, 
and define elementary screens and screen networks.

A causal set $\C$ is a locally finite, partially ordered set of events.
That is, if we denote the events by $p,q,r,\ldots$ and,
say, $p$ precedes $q$, we write $p\leq q$. The equal option is used when 
$p$ coincides with $q$.  
The causal relation is reflexive, i.e.\ $p\leq p$ for every event $p$ 
and transitive, i.e.\ if $p\leq q$ and 
$q\leq r$, then $p\leq r$.  It is also antisymmetric, that is, 
if $p\leq q$ and $q\leq p$, then $p=q$, which ensures that there are 
no closed timelike loops in the causal set.  Local finiteness means 
that, given $p\leq q$, there is a finite number of events which are both in the 
future of $p$ and in the past of $q$. 

A causal relation $p\leq q$ is called an ``edge'' or a
``covering relation'' if it is not implied by transitivity from other
relations in the history.  
      
In our previous work, we have used the term ``causal history'' to 
describe causal sets whose events carry additional structure. 
For example, in causal histories of spin 
networks \cite{FM2,tubes,pqtubes} the 
events are local changes in spin networks, while in 
\cite{pqtubes} they are local changes in $(p,q)$ 
string networks.  
For the reasons we explained in the introduction, in this paper we
are concerned with histories whose events are {\it 
elementary screens}.
An elementary screen $s$ is a quadruple
    \f
    s\equiv  \left \{  
    s_{L}^{-} ,s_{R}^{-} ,s_{L}^{+} , s_{R}^{+} \right \}. 
    \label{eq:screen}
    \ff
It has four components, the past left, $s_{L}^{-}$,
past right, $s_{R}^{-}$, future left, $s_{L}^{+}$ 
and future right $s_{R}^{+}$.  Within each quadruple there are 
maps from the past left component to the future right one, and from
the past right to the future left, namely, 
   \begin{eqnarray}
       \LR: s_{L}^{-} & \rightarrow & s_{R}^{+}, \nonumber \\
       \RL: s_{R}^{-} & \rightarrow & s_{L}^{+}.  \nonumber \\
       \label{eq:screenmaps}
   \end{eqnarray}
These maps are the contribution of the screen $s$ to the dynamics 
of the causal history.  

A special case of a causal history is when all the events are 
elementary screens. We will call such a history a {\it 
screen network}.  It is a partially ordered set of screens, in 
which two screens are related, $s\leq t$, when one of the future 
components of $s$ precedes one of the past components of $t$.  The 
following condition is imposed on a screen network:  There can be at 
most one edge (covering relation) from $s$ to $t$.  This means that, 
if $s$ is in the immediate past of $t$, $t$ can only ``see'' one side 
of $s$.  

Thus, the network describes signals
exchanged amongst a set of elementary surfaces that make up 
a quantum version of a spacetime.  
We have called the two sides of a 
screen $L$ and $R$, and each has a future and a past. 
According to (\ref{eq:screenmaps}),
information that comes into  the past of the left side of a screen in 
the network may exit
only from the right side of a screen, and vice versa. Thus, information is 
carried from one side of the screen to the other by the internal 
maps $\LR$ and $\RL$.  Components of different screens are related by 
the external (to the screen) maps $\leq$.  These relations 
encode the causal structure of the screen network.

Having defined the screen network, the following sets can be 
constructed from it and will be used in the remaining of this paper. 

\begin{itemize}
    
\item{}The causal past, $\CP(s)$, of a screen $s$ is the set of screens 
$t$ in the screen network with $t\leq s$.

\item{}The left null past, $\LNP$, of a component of a screen $s$ is 
the set of those screens in its causal past that are related to $s$ 
by a sequence of alternating  $\LR$ and $\leq$ maps.  Its right null 
past, $\RNP$, is the set of those past screens that are related to $s$ 
by a sequence of alternating  $\RL$ and $\leq$ maps.  The null past 
of $s$ is the union of these two sets.  

\item{}The timelike past $\TP(s)$ of a screen $s$ is the screens in its causal 
past that are not null related to $s$, that is,
${\TP}(s)  = {\CP}(s)- \NP(s)$.

\item{}The causal future $\F(s)$, null future $\NF(s)$  and timelike 
future $\TF(s)$ of $s$ are  similarly defined.  

\end{itemize}

It is interesting to note that the two sides of a screen 
(the two internal maps) allow this natural  distinction between 
null and timelike.  Of course, for a general screen network,  
there is no global decomposition into left and right flows of 
information.  Still, the two-sided nature of the screens gives the 
elementary processes and the flow of information in the network
a chiral aspect, as left and right flows can always be distinguished
locally.

A screen network can be reduced to 
its underlying causal set by removing the internal $\LR$ and $\RL$ 
maps and compressing the four screen components to a single causal set event.

\section{The quantum screen network}

We next wish to turn a screen network $\Screen$ into a network of 
elementary quantum-mechanical systems.  In doing so, we will assume 
that quantum information propagates without change between screens 
and undergoes non-trivial evolution only when going through a screen.  
We express this by assigning a Hilbert space to every edge of 
the screen network, and two (unitary) evolution 
operators to each screen.\footnote{
Another possibility is to do the reverse, namely, turn each 
screen into a finite-dimensional Hilbert space and each edge
into an evolution map.  However, as it was discussed in 
\cite{FM4}, as soon as this is done, acausal evolution becomes possible 
and the quantum mechanical information flow does not reflect the 
underlying causal set anymore.   The solution that was 
proposed in \cite{FM4} is the recipe used here, i.e.\ 
attach the Hilbert spaces on the edges
and the operators to the events.  This is well-motivated 
physically as it agrees with the intuition that events should 
represent change, and so their quantum-mechanical counterpart should 
be an operator rather than a state space. }

Before we give the definition of the quantum screen 
network\footnote{We may note that this differs from the 
quantum causal sets defined by 
Criscuolo and Waelbroeck\cite{CW}.}, we  
list the two of its desired features that serve as the starting point. 
First, we wish to replicate the fact that, in quantum mechanics, the 
composite state space of spacelike separated systems is the tensor 
product of the individual state spaces.  The individual systems in 
the screen network case are the edges $e_{i}$ connecting different 
screens.  Each edge is represented by a finite-dimensional state space 
$H(e_{i})$.  Two such edges are spacelike separated when there is no 
null or timelike path from the one to the other.  Thus, given a set 
of spacelike separated edges, $a=\{e_{1},e_{2},\ldots,e_{n}\}$,  the 
composite state space is $H(a)=H(e_{1})\otimes H(e_{2})\otimes\ldots
\otimes H(e_{n})$. 

Second, if such a set of edges $a$ is in the past of another set $b$, 
we can only expect to have a unitary evolution map from $H(a)$ to 
$H(b)$ if there has been no ``loss'' or ``gain'' of information from 
$a$ to $b$.  What this means for a screen network is the following. 
Consider two edge-sets $a$ and $b$ containing no common edges.
Let  every edge in $a$ be in the past of some edge in $b$.  
Furthermore, let every edge in $b$ be in the future of some edge in $a$.
Then, in the notation of \cite{FM4}, $a$ and $b$ are a 
{\it complete pair}.   Since there are no edges in the future of $a$ 
that are spacelike to $b$ and no edges in the past of $b$ that are 
spacelike to $a$, a complete pair serves as a model of information 
conservation.  In a quantum screen network, we expect to have 
unitary evolution only between edge sets that are complete pairs.  

Keeping the above in mind, we define the {\it edge screen network}, 
$\ES$, to be the partially ordered set whose elements are edge-sets, 
sets of spacelike separated edges $a, b,\ldots$ in the screen 
network $\Screen$.  Two edge-sets are related in $\ES$ when they are 
a complete pair.  

We  may now define the quantum screen network as a functor from the 
edge screen network to the category of Hilbert spaces (which has 
Hilbert spaces for its objects and unitary operators as its arrows). 
Hence, a quantum screen network $\QSN$, is the functor
\f
\QSN:\ES \rightarrow Hilb,
\label{functor}
\ff
such that for every edge-set $a$ in $\ES$ there is a 
finite-dimensional Hilbert space $H(a)$ in 
$\QSN$. If $a$ and $a'$ are spacelike separated (have no common edges), 
$H(a\cup a')=H(a)\otimes H(a')$.
For every complete pair $a\leq b$ in $\ES$, $\dimension 
H(a)=\dimension H(b)$, and there is a unitary evolution operator 
$E:H(a)\longrightarrow H(b)$ in $\QSN$.

According to the above, for some screen $s$ in the screen network, 
$H(s_{L}^{-})$ is the state space of the edges going into the left of 
the screen, and $H(s_{R}^{-})$ the state space of the edges going 
into the right of the screen.  A screen has the same information 
capacity on both sides, which implies that 
\beq
\dimension H(s_{L}^{-})=\dimension H(s_{R}^{-}).
\label{equal}
\eeq
 
Clearly, $s_{L}^{-}$ and $s_{R}^{+}$ is a complete pair, and so is  
$s_{R}^{-}$ and $s_{L}^{+}$.  The unitary operators in $\QSN$ 
corresponding to these two complete pairs are $\widehat{\LR}(s)$ and
$\widehat{\RL}(s)$.  By the unitarity of the operators in $\QSN$
we have
\beq
\dimension H(s_{L}^{-})=\dimension H(s_{R}^{+})   \ \ \mbox{and} \ \ 
\dimension H(s_{R}^{-})=\dimension H(s_{L}^{+}).
\eeq
Thus, the state 
spaces of all the components of a screen $s$ have the same dimension, 
which we denote $D(s)$.

We define the area of a screen $s$ 
to be a measure of the information capacity of a 
screen. This is proportional to the dimension of the Hilbert space 
of any of the components of $s$:
\f
A(s) \equiv a l_{Planck}^2 \ln D(s).
\label{area}
\ff
where $a$ is a constant, which we may take to
be equal to $1/4$ to agree with the semiclassical Bekenstein bound.
Finally, any evolution operator $E_{ab}:H(a)\rightarrow H(b)$ 
in $\QSN$ can be decomposed into 
$\widehat{\RL}$ and $\widehat{\LR}$ operators in the screens between 
$a$ and $b$.  

We claim that a quantum screen network is a holographic theory 
because the generating evolution operators $\widehat{RL}$ and 
$\widehat{LR}$ act on Hilbert spaces on one side of some screen. Simply 
because a screen has two sides, we regard it as the quantum spacetime 
analogue of a spacetime object of codimension 2. 

\section{Causal spin network evolution}

We turn now to the question of 
what restrictions may be imposed on  candidates for
background independent quantum theories of gravity by
the requirement that the corresponding quantum
causal histories are screen networks.  
We consider here the  example of causal histories of 
spin networks, in the original form given in  \cite{FM2},
or the extensions in \cite{tubes,pqtubes,mpaper}.  
We first review why the histories in such theories are
quantum causal histories\cite{FM4},
and then ask what additional conditions must be satisfied to 
ensure that they are screen networks.  

In these histories, the analogue of a 
spatial region in a spacetime is an open spin network 
$\gamma$.  This is an 
oriented graph (or, in \cite{tubes,pqtubes,mpaper}, 
a punctured two-dimensional surface) 
with free ends whose edges are labeled by
representations of a quantum group or supergroup.  In the case
of quantum general relativity this is taken to be 
$SU_{q}(2)$. Extensions to supergravity\cite{yilee1} 
or other dimensions are
described by different quantum groups.  We denote  
the labelled edges by $e_{i},e_{j}$, etc.

An important observation is that 
$\gamma$  labels a state in the space of 
intertwiners ${\cal V}_{\{e_{i}\}}$ of the representations labeling 
its free edges.
The dimension of ${\cal V}_{\{e_{i}\}}$, given the labels on the free edges, 
can be calculated using the Verlinde formula in \cite{verlinde}.

The open labelled graph $\gamma$  is generally a piece of a closed spin 
network $\Gamma$  
which defines the quantum geometry of a complete spacelike
slice of a spacetime history. See \cite{FM2,tubes} for details.

A local evolution move replaces $\gamma$ with a new open graph, 
$\gamma^{\prime}$, which has the same free ends $\{e_{i}\}$ as $\gamma$.
The result is a bubble evolution move, in which only the
local region $\gamma$ evolves, leaving unchanged the 
remaining $(\Gamma -\gamma)$.  As a result, the history is 
a quantum version of many-fingered time evolution.

By construction, $\gamma$ and $\gamma^{\prime}$ have the same 
labeled free edges and therefore live in the same space of intertwiners
${\cal V}_{\{e_{i}\}}$.  The move which replaces $\gamma$ by
$\gamma^{\prime}$ is then represented by a transition in the
Hilbert space, ${\cal V}_{\{e_{i}\}}$.  The dynamics of the theory can then 
be given by a rule which assigns  
an evolution operator to each such space of intertwiners. In this 
case,
\beq
\widehat{E}: {\cal V}_{\{e_{i}\}} \longrightarrow
         {\cal V}_{\{e_{i}\}}.
	\label{evolve}
\eeq
To ensure that there is no loss of information {\it locally},
these operators are required to be unitary.  All the generating 
evolution moves listed in \cite{FM2}, the so-called Pachner moves 
for abstract spin networks, are operators of this type.  

This shows that to each causal spin network history $\cal M$,
as described in \cite{FM2,tubes,pqtubes,mpaper}, 
there corresponds a quantum causal
history $Q{\cal M}$.  Each open spin network piece $\gamma$ in ${\cal 
M}$ is a Hilbert space ${\cal V}_{\{e_{i}\}}$ in $Q{\cal M}$.  Every 
time there is an evolution move $\gamma$ to $\gamma'$ in $\cal M$, 
$\gamma$ and $\gamma'$ are a complete pair.  $\gamma'$ lives in the 
same Hilbert space ${\cal V}_{\{e_{i}\}}$ as $\gamma$ and thus the 
move is a unitary operator in $Q{\cal M}$.

Now we turn to the additional requirements that arise if we want
each such move to correspond to an elementary screen.  This
requires that we do two things.  First, in each transition
it must be possible to pick out four sets of
edges, corresponding to the four components of a screen. Second,
Hilbert spaces must be associated to them in such a way that the
evolution splits into two parts according to eq.\ (\ref{eq:screenmaps}).

To accomplish this  we note that the space of intertwiners
${\cal V}_{\{e_{i}\}}$ can be split as follows.  We divide the
external edges $e_{i}$ into two sets, which we will call the
left set $e_{L}$ and the right set $e_{R}$.  We may then write
\f
{\cal V}_{\{e_{i}\}}=\bigoplus_{j}{\cal V}_{\{e_{L}\}j} 
                     \otimes {\cal V}_{\bar{j}\{e_{R}\}}
\ff
where $j$ is short for $e_{j}$,
the the sum is over a complete set $j$ of the representations
and $\bar{j}$ is the complex conjugate representation.

We require that this choice be made  so that for at least
one $j$,
\f
\dimension {\cal V}_{\{e_{L}\}j} = \dimension {\cal 
V}_{\bar{j}\{e_{R}\}} .
\label{eq:restriction1}
\ff
We then pick a particular $j=j_{0}$ that satisfies this 
and restrict $\gamma$ and 
$\gamma^{\prime}$ to lie in the subspace
${\cal V}_{\{e_{L}\}j_{0}} \otimes {\cal V}_{\bar{j}_{0}\{e_{R}\}} $.
The evolution operator $\widehat{E}$ is then required to be of the form,
\f
\widehat{E}=  \left ( 
\begin{array}{cc}
    0 & \widehat{LR}  \\
    \widehat{RL} & 0
\end{array} \right). 
\label{eq:restriction2}
\ff
This agrees with the general form implied by the application of
eq.(\ref{functor}) to   eq.(\ref{eq:screenmaps}) if 
\f
H(s_{L}^{\pm}) ={\cal V}_{\{e_{L}\}j_{0}}  
\ff
and
\f
 H(s_{R}^{\pm}) ={\cal V}_{\{e_{R}\}\bar{j}_{0}}   . 
\ff
The two restrictions (\ref{eq:restriction1}) and (\ref{eq:restriction2}) are
non-trivial, so most causally evolving spin network histories
are not screen networks.  But it is not difficult to construct
examples  that do satisfy the conditions.  For example, for
$SU_{q}(2)$ spin networks we
may restrict all labels to spin $1$ and all nodes to be four
valent, then the splitting may be accomplished with $j_{0}=1$
at all transitions.
Finally, we note that since there is no requirement that
$\widehat{LR}=\widehat{RL}$ the resulting theory may be chiral.

\section{The weak holographic principle}

It is not difficult to abstract from the definition of a quantum 
screen network the main elements of what we propose make a theory 
holographic, and formulate them as the {\it weak holographic principle}.

These are:
\begin{enumerate}
    \item
    A discrete holographic theory is based on a causal 
    history, that is,  the events in the quantum spacetime form a 
    partially ordered set under their causal relations. 
    \item
    Among the elements of the quantum spacetime, a set of 
    screens can be identified.  Screens are 2-sided objects with
    two past sides and two future sides.  
    \item
    There is a Hilbert space for each past or future side of a screen.
    Observables on this Hilbert space describe information that an 
    observer at the screen may acquire about the causal past of the 
    screen, by measurements of fields on that side of 
    the immediate past of the screen.  There is an algebra of such 
    observables for each side of the screen.  
    \item
    Since screens are 2-sided, each has an orientation reversal 
    operation that sends the state space of one side to its complex 
    conjugate on the other side.
    \item
    All observables in the theory are operators in the algebra of 
    observables ${\cal A}(s)$ for some screen $s$. 
    \item
    The area of a screen $s$ is either a fixed number $a_{s}$, or an 
    operator  $\widehat{A}_{s}$ in ${\cal A}(s)$.  
    If it is a number, it is proportional 
    to the dimension $D(s)$ of the Hilbert space of either screen side,
    \beq
    a_{s}\propto l_{Planck}^{2}\ln D(s).
    \label{eq:yes}
    \eeq    If it is an operator,
    \f
    {\cal H}_{s}=\bigoplus_{a} {\cal H}_{s}^{a}
    \ff
    where each factor ${\cal H}_{s}^{a}$ is the eigenspace of
    $\hat{A}[{s}]$ with eigenvalue $a[{s}]$ which each
    satisfy (\ref{eq:yes}).
\end{enumerate}

\section{Conclusions}

In this paper, we listed and analyzed the main features that can be 
expected of a holographic theory of quantum cosmology.  Based on this, we
stated the holographic principle in a discrete, background independent 
form. 

More can be said about the relationship of the weak holographic 
principle to its ``strong'' forms given 
elsewhere \cite{thooft-holo,lenny-holo,willylenny,RB}.  
As we mentioned in the introduction, what needs to be checked is 
that, when the weak holographic history has a good continuum limit, 
the strong holographic principle holds in this limit continuum 
theory.  At this stage,  little is known about the continuum limit of 
discrete causal theories like (quantum) screen 
networks\cite{fmls1,withstu}.  Ambjorn,
Loll, Anagnostopoulos and others have shown that Lorentzian 
$1+1$ gravity belongs to a universality class different that 
Liouville gravity \cite{AL,ANRL}.  
Similar calculations in higher dimensions are 
technically very demanding, and it is expected that the results in the 
causal/Lorentzian case are very different than the euclidean ones. 

Requiring that a theory is weakly holographic places constraints on 
both its dynamics and the algebras of observables.  The way it affects 
the measurement theory that is appropriate in background independent 
theories of quantum cosmology will be discussed in \cite{fmls4}.
 
Before closing, we briefly consider the possibility that there be
a version of the strong holographic principle that may hold at the
background independent level.  Given the formalism developed here
it is possible to state a strong form of the holographic
principle for a background independent theory.  This makes it
possible to identify a problem that would have to be overcome to
realize it in the kinds of theories we have
considered here.   

The strong form of the holographic principle has been stated
in certain backgrounds, such as $AdS/CFT$ as a conjecture about
an equivalence between a boundary theory and a bulk theory\cite{AdS}.
There is no boundary in a cosmological theory, but a screen
network such as defined here plays the role of the boundary
theory as it describes evolution in terms of a flow of information
between Hilbert spaces attached to screens.  
 
A strong form of the holographic principle then requires that there be a bulk
theory  which has the property that its kinematics and dynamics
is exactly equivalent to some screen network theory.  
Since the theory is expected to be cosmological, so that there
is no boundary, the screens are
embedded in the bulk. This means that the bulk theory must have
the property that it is equivalent to a different theory that involves
only a subset of events which are its screens.

It is easy to imagine that there is a sense in which a
sub-history of any history  may be defined which gives an approximate
description of that history. This could be accomplished by an
appropriate coarse-graining.  But the strong holographic principle 
requires more, the screen theory must not just arise from a 
coarse-graining of the bulk theory, it must be completely equivalent to 
it.  

We can formulate this in the class of theories considered
here.  For the bulk theory, we consider a general quantum causal 
history $Q\C$.
We know from the above that this includes some candidates for 
background independent quantum theories of gravity.
The strong holographic principle would state that for every
$Q\C$ satisfying a certain list of conditions, there is a quantum
screen network $\QSN$ which is equivalent to it.  Equivalence
requires that the relationship be $1$-to-$1$ so that $Q\C$ 
can be recovered from $\QSN$.  

While this is not impossible, it is not 
difficult to see what kinds of obstacles would have to be
overcome to accomplish it.  To do this we consider how the
results of this paper may be extended to the case of a general
quantum causal history, some of whose events satisfy the conditions
to be screens.  
Let then $Q\C$ be a quantum causal history as described in
\cite{FM4}.  Namely, if $\C$ is the underlying causal set, $Q\C$ is the 
functor  
\f
{Q\C} : E\C\rightarrow Hilb
\ff
where $E\C$ is the poset of edge-sets of $\C$, in which two 
edge-sets are related when they form a complete pair as defined in section 
3.  We call a causal history complete when there is an {\it initial} 
edge-set $A_{0}$ such that $\mbox{\sl Future}(A_{0})=\C$.  An initial 
state is a choice of $|\Psi\rangle \in H(A_{0})$.  Given $Q\C$,
each such initial 
state determines a density matrix in the Hilbert space of any other 
edge-set in the causal history.   It follows that, if a quantum causal 
history contains quadruples of events which are screens, a choice of 
initial state will determine a density matrix in each screen Hilbert 
space, corresponding to information flowing across the screen.  

For the strong holographic principle to hold in the form just stated, 
two things are required.  First, it must be possible to find a screen 
network containing the subset of events of $\C$ that are screens, and 
with the property that the density matrix in any screen Hilbert space 
is fully determined by the density matrices on its past screens.  
Second, it should be possible to reconstruct $Q\C$ from $\QSN$.  

Even if the first can be done, there is a general difficulty with the 
second.  The problem is that  there is no
natural notion of a quantum causal history which is a subhistory
of another. Where one causal set may be a subset
of another, the same is not the case for the corresponding 
quantum causal histories, as the covering relations are singled out 
in the construction. Since not all covering relations of the 
screen subset are covering relations in the causal set,
there is no natural restriction of the functor $Q\C$ that reduces
the quantum causal history to a quantum causal history on the subset.

This is related to the fact that the flow of information through a quantum
causal history is path-dependent,  which is also 
a feature of the flow of information in the semiclassical
theory on a general, curved, spacetime.  The quantum information that flows
between two screens will in general be a superposition of the effects
of several evolution operators.  

Since there are always 
fewer screens in $\Screen$ than events in the original $\C$, the 
number of covering relations in $\Screen$ are less than those in 
$\C$.  Thus, it will not, in general, be possible to invert the 
procedure and use the data in $\QSN$ to determine a unique 
quantum causal history $Q\C$.  Rather, a definition of a subhistory 
of $Q\C$ should involve a suitable notion of coarse-graining in which 
information about the original history is lost.  Unless this can be 
avoided, there will not be a unique bulk history $Q\C$ which is 
determined from the screen network history $\QSN$.

In either case, what will be true is that the theory will
satisfy the weak holographic principle, in the form we have
given it here.

\section*{Acknowledgments}

We would like to thank Raphael Bousso, Louis Crane, 
Sameer Gupta, 
Chris Isham, Ted Jacobson, Yi Ling, Mike Reisenberger and Carlo Rovelli 
for useful discussions and correspondence.  LS thanks in addition
Willy Fischler and Reza Tavakol for stimulating discussions. We are
also grateful to David Gross and Jim Hartle for hospitality at
the Institure for Theoretical Physics and Physics Department at UCSB,
where this work was begun.  
This work was supported by NSF grants PHY/9514240 and PHY/9423950
to the Pennsylvania State University and a gift from the Jesse 
Phillips Foundation.

\end{document}